\documentclass[conference]{IEEEtran}
\IEEEoverridecommandlockouts
\usepackage{cite}
\usepackage{amsmath,amssymb,amsfonts}
\usepackage{algorithmic}
\usepackage{graphicx}
\usepackage{textcomp}
\usepackage{xcolor}

\usepackage{tabularx}
\usepackage{multirow}
\usepackage[belowskip=-5pt,aboveskip=0pt]{caption}
\usepackage{array}
\newcolumntype{P}[1]{>{\centering\arraybackslash}p{#1}}
\newcolumntype{M}[1]{>{\centering\arraybackslash}m{#1}}
\newcommand{\PreserveBackslash}[1]{\let\temp=\\#1\let\\=\temp}
\newcolumntype{R}[1]{>{\PreserveBackslash\raggedleft}p{#1}}
\usepackage{tabulary,booktabs}
\usepackage{lipsum}
\usepackage{blindtext}
\usepackage{amsmath,amssymb}
\usepackage{hyperref}

\def\BibTeX{{\rm B\kern-.05em{\sc i\kern-.025em b}\kern-.08em
    T\kern-.1667em\lower.7ex\hbox{E}\kern-.125emX}}
\begin{document}

\title{Multiscale deep context modeling for lossless point cloud geometry compression}

\author{\IEEEauthorblockN{Dat Thanh Nguyen, Maurice Quach, Giuseppe Valenzise, Pierre Duhamel}
\IEEEauthorblockA{\textit{Université Paris-Saclay, CNRS, CentraleSupelec, Laboratoire des Signaux et Systèmes} \\
91190 Gif-sur-Yvette, France\\
{\{thanh-dat.nguyen, maurice.quach, giuseppe.valenzise, pierre.duhamel\}@l2s.centralesupelec.fr}
}}

\maketitle

\begin{abstract}
We propose a practical deep generative approach for lossless point cloud geometry compression, called MSVoxelDNN, and show that it significantly reduces the rate compared to the MPEG G-PCC codec. Our previous work based on autoregressive models (VoxelDNN~\cite{nguyen2020learning}) has a fast training phase, however, inference is slow as the occupancy probabilities are predicted sequentially, voxel by voxel. In this work, we employ a multiscale architecture which models voxel occupancy in coarse-to-fine order. At each scale, MSVoxelDNN divides voxels into eight conditionally independent groups, thus requiring a single network evaluation per group instead of one per voxel. We evaluate the performance of MSVoxelDNN on a set of point clouds from Microsoft Voxelized Upper Bodies (MVUB) and MPEG, showing that the current method speeds up encoding/decoding times significantly compared to the previous VoxelDNN, while having average rate saving over G-PCC of 17.5\%. The implementation is available at \url{https://github.com/Weafre/MSVoxelDNN}.

\end{abstract}


\section{Introduction}
\par A point cloud is a set of 3D points, where each point is associated to spatial coordinates $x, y, z$ (geometry) and attributes (color, reflectance, etc.). Unlike 2D images, the irregular spatial sampling of point clouds makes the coding task more  challenging than for traditional video. As point clouds are the a commonly used data structure in many applications such as immersive communication, autonomous vehicles, cultural heritage, etc., efficient compression methods are required for enabling effective point cloud transmission/storage.
\par Two Point Cloud Compression (PCC) standards have been developed by the Moving Picture Expert Group (MPEG)~\cite{graziosi2020overview}: Video-base PCC (V-PCC) and Geometry-based PCC (G-PCC). V-PCC is based on 3D-to-2D projections and the 2D image/video compression standards are utilized to encode the projected data. On the other hand,  G-PCC processes point clouds directly in the 3D space. The geometry and attribute information of a point cloud are first separated and G-PCC encodes them independently. Prior to actual geometry coding, the spatial coordinates of the points are first quantized to integer precision (voxelization). Once the PC is voxelized, its geometry can be represented in the voxel domain or octree domain. In particular, a binary occupancy map is defined over the voxel grid to indicate whether a voxel contains at least one point. This is the signal that we aim at coding.

\begin{figure}[tb]
\centering
\includegraphics[width=0.85\linewidth]{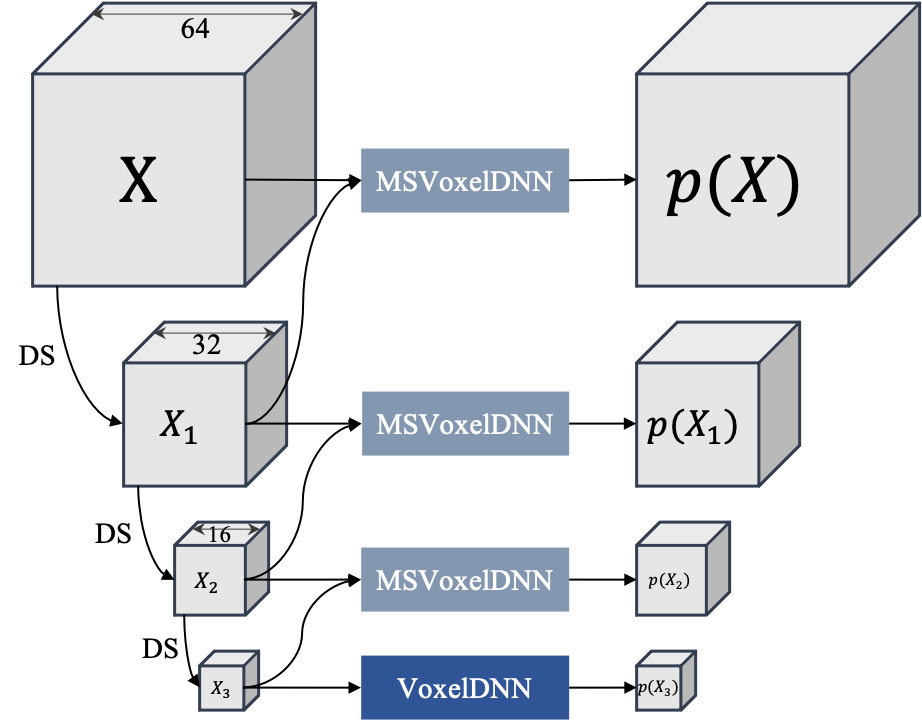}
\vspace{0.1cm}
\caption{Overview of the MSVoxelDNN architecture with input block of size 64 and 3 scales. DS is the downsampling operation (max-pooling). The base resolution of size $8$ is encoded using a VoxelDNN context model. The higher resolutions are predicted from lower resolution as well as encoded groups at the same scale. The predicted block probabilities on the right side are passed to an arithmetic coder for encoding voxel occupancies. The final bitstream is the concatenation of all bits at all scales.}
\label{fig:msvoxeldnnarchitecture}
\end{figure}

\par In order to efficiently code the PC geometry losslessly, it is necessary to accurately estimate the occupancy probabilities to be employed into a context adaptive arithmetic codec. In our previous work, we have modeled the voxel occupancy distributions using a likelihood-based deep autoregressive network called VoxelDNN \cite{nguyen2020learning}, inspired by the popular PixelCNN model~\cite{van2016pixel}. VoxelDNN achieves state-of-the-art gains (up to 34\%) over the MPEG G-PCC reference codec. 

Autoregressive models can accurately predict probability distributions. However, the decoding process using this approach is equivalent to sampling from the high-dimensional conditional distribution of voxel occupancies, which is computationally complex as it demands one network evaluation per voxel. 
In this work, taking inspiration from previous work in 2D image generation~\cite{reed2017parallel}, we propose a multiscale method (named MSVoxelDNN) for lossless geometry compression of static dense point clouds which addresses the complexity problem of VoxelDNN. Our main contributions are:  
\begin{itemize}
    \item We introduce for the first time a multiscale deep context model in the voxel domain to estimate occupancy probabilities, in which higher-resolution scales are modeled conditioned on the lower-resolution ones. 
    \item We accelerate the inference by parallelizing voxel prediction. At each scale, voxels are partitioned into groups (see Figure~\ref{fig:groupdependencies}). Voxels belonging to the same group are assumed to be conditionally independent from each other. In this way, we can predict all the voxels of the same group \textit{simultaneously}, reducing the computation time. Instead, each group of voxels is assumed to depend on the previously decoded ones, and thus out context model can leverage dependencies between groups.
    
\end{itemize}

\par Compared to VoxelDNN, we make an approximation in that we do not utilize the statistical dependencies of voxels inside groups (due to the conditional independence assumption). We demonstrate experimentally that this approximation entails only a small loss of performance compared to the original VoxelDNN, and still outperforms significantly MPEG G-PCC in terms of bits per occupied voxel. However, in terms of complexity, MSVoxelDNN is on average 35 and 109 times faster compared to VoxelDNN for encoding and decoding, respectively. The rest of the paper is structured as follows: Section \ref{relatedwork} reviews related work; the proposed MSVoxelDNN method is described in Section \ref{method}; Section \ref{experiment} presents the experimental results; and conclusions are drawn in Section \ref{conclude}.


\section{Related Work}

\par To deal with the irregular distribution of points in 3D space, many PCC methods employ octree representations \cite{schnabel2006octree,7434610,6224647,garcia2017context,garcia2018intra,garcia2019geometry,huang2020octsqueeze} or local approximations \cite{dricot2019adaptive}. The octree based method P(PNI) proposed in \cite{garcia2019geometry} builds a reference octree using an intra prediction mode. Each octant is then encoded with 255 contexts and a 255$\times$255 frequency table must be transmitted to the decoder. In the MPEG G-PCC codec, geometry can be represented by a pruned octree plus a surface model (trisoup coder) or a full octree (octree coder). To exploit local geometry information within the octree and obtain an accurate context for arithmetic coding, the G-PCC octree coder introduces many techniques such as Neighbour-Dependent Entropy Context \cite{neighbor}, intra prediction \cite{intracodinggpcc}, planar/angular coding mode \cite{planarcodingmode, angularcodingmode}, etc. Instead, in this paper, we represent the PC geometry in a \textit{hybrid} mode, mixing the octree and voxel domains. On the one hand, octree can adapt to the sparsity of the point cloud, as partitioning stops at the empty node; on the other hand, geometric information are kept and can be naturally processed by a neural network.

\par Recently, deep learning has been applied widely in point cloud coding in both the octree domain \cite{huang2020octsqueeze,biswas2020muscle} and especially voxel domain \cite{8954537,quach2019learning,wang2019learned, nguyen2020learning}. A coding method for static LiDAR point cloud is proposed in \cite{huang2020octsqueeze} which learns the probability distributions of the octree based on contextual information and uses an arithmetic coder for lossless coding. In this work we focus instead on \textit{dense} point clouds, where voxel-based approaches have shown interesting results. In particular, our recent work, VoxelDNN~\cite{nguyen2020learning}, is an auto-regressive based model which predicts the distribution of each voxel conditioned on the previously decoded voxels. VoxelDNN obtains an average rate saving of 30\% over G-PCC. The auto-regressive approach of VoxelDNN is similar to PixelCNN \cite{van2016pixel} which provides accurate 2D data likelihood estimations. However, the common problem of auto-regressive models is the complexity, as these models require a network evaluation per voxel/sub-pixel. In 2D, several methods have been proposed to overcome this limitation. PixelCNN++ \cite{salimans2017pixelcnn++} models the joint distribution of three color channels simultaneously and proposes several optimizations to PixelCNN. Multiscale PixelCNN \cite{reed2017parallel} generate pixels in certain groups in parallel. The L3C method \cite{mentzer2019practical} employs a latent space to facilitate the learning of conditional probability estimates at several scales. While this technique solves the \textit{complexity} issue, the estimated probabilities are not accurate enough and the coding gains are significantly less interesting than using Pixel CNN. In this paper, we aim at finding a good trade-off between complexity and compression performance. Thus, we follow the principle of~\cite{reed2017parallel} introducing parallelization and multiscale prediction.
\label{relatedwork}

\section{Proposed Method}

\begin{figure}[tb]
\centering
\includegraphics[width=0.99\linewidth]{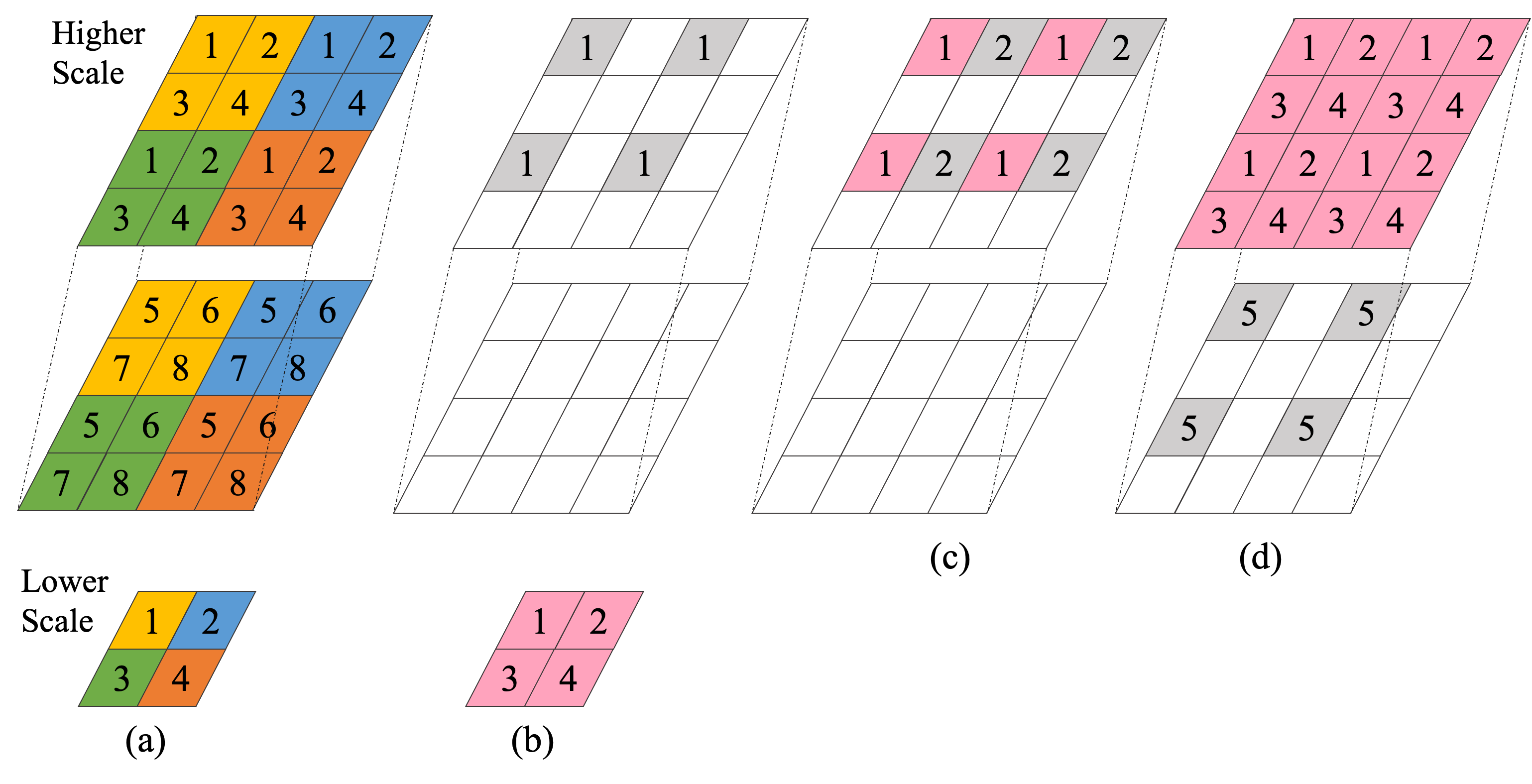}
\vspace{0.1cm}
\caption{Prediction parallelization in MSVoxelDNN/ (a) partitioning of a block into groups of conditionally independent voxels. For the sake of clarity and without loss of generality, we show the context modeling for a block of size $2 \times 4 \times 4$. Downsampling is achieved by applying a $2\times 2 \times 2$ max pooling operator, i.e., the voxels in the lower scale are the maximum of all voxels having the same color on higher scale (MaxPooling operation). (b), (c) and (d) illustrate some steps of the groups prediction. The target groups are in gray, while the input (context) groups are in pink. (b): the $1^{st}$ group is predicted from all the groups at the lower resolution. (c) the $2^{nd}$ group is predicted from group 1 at the same resolution. (d) the $5^{th}$ group is predicted from group 1,2,3 and 4 (at the same scale).}
\label{fig:groupdependencies}
\end{figure}

\begin{figure*}[tb]
\centering
\includegraphics[width=0.87\linewidth]{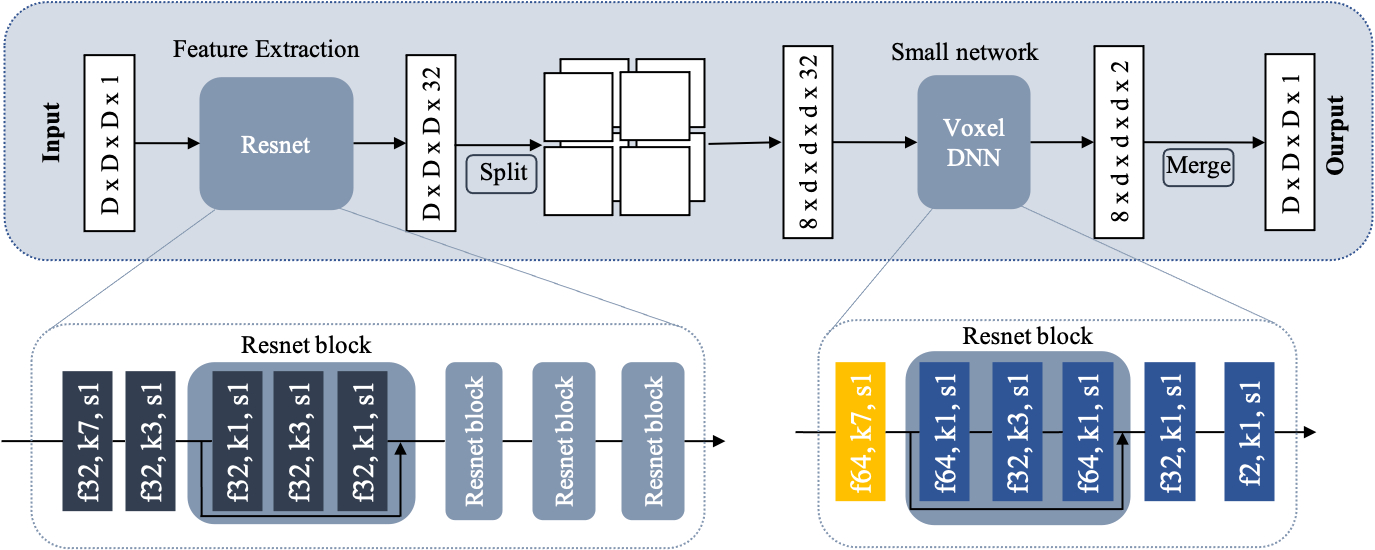}
\vspace{0.1cm}

\caption{ Group prediction network architecture. This network predict group 2 from group 1, corresponding to (c) in the example in Figure~\ref{fig:groupdependencies}. The only learnable modules are Resnet and VoxelDNN. Merge and split operations only reshape data. We use a sequence of Resnet blocks to extract features from input. The features are then spatially split into smaller blocks before parallel processing by a small VoxelDNN. Black rectangular blocks are normal 3D convolution with `f32,k7,s1' stands for 32 filters, kernel size 7 and stride 1. All convolutional blocks of VoxelDNN are masked convolutions \cite{nguyen2020learning}, type A mask is in the first layer (in yellow), followed by type B masks. }
\label{fig:networkarchitecture}
\end{figure*}

\par As mentioned before, in this paper, we focus on voxelized point clouds. Without loss of generality, we assume the point cloud contains $2^n \times 2^n \times 2^n $ voxels. An octree is obtained by recursively spliting the voxel volume into eight sub-cubes until the desired precision is achieved. The occupied cube is marked by bit 1 and empty cube is marked by bit 0. As a result, in each octree node, the generated 8 bits represent the occupancy of the 8 child nodes. A point cloud of size $2^n \times 2^n \times 2^n $ can be represented by an $n$ level octree. In this work, and similar to~\cite{nguyen2020learning}, we partition an $n$-depth point cloud up to level $n-6$, and thus obtain a $n-6$ high level octree and a number of non-empty binary blocks $v$ of size $2^6 \times 2^6 \times 2^6$, which we refer to as resolution $d=64$. The high-level octree allows to coarsely remove most of the empty space in the point cloud, which does not contain any useful context information to predict occupancies. All the non-empty voxel blocks are further processed with our multiscale scheme. We first define a raster scan order in the 3D space that scans one voxel at a time in depth, height and width order.  We index all voxels in block $v$ at resolution $d$ from 1 to $d^3$ in 3D raster scan order with:
 \begin{equation}
    v_i= 
    \begin{cases} 
    1, \quad \text{if $i^{th}$ voxel is occupied}\\
    0, \quad \text{otherwise}.
    \end{cases}
\label{definition}
\end{equation}

\subsection{VoxelDNN context model}
VoxelDNN~\cite{nguyen2020learning} factorizes the joint distribution of a voxel block into a product of conditional distributions: 
\begin{equation}
    p(v)= \underset{i=1 }{\overset{d^3}{\Pi}}p(v_i|v_{i-1},v_{i-2},\ldots,v_{1}).
    \label{eq:p(v)}
\end{equation}
Each term $p(v_i|v_{i-1}, \ldots, v_1)$ above is the occupied probability of the voxel $v_{i}$ given only the occupancy of previous voxels,  referred to as \textit{causality constraint}. All factors in equation~\eqref{eq:p(v)} are estimated by a neural network with masked filters to enforce causality~\cite{nguyen2020learning}. Therefore, the inference must also proceed sequentially voxel-by-voxel and VoxelDNN performs one network evaluation per voxel.

\subsection{MSVoxelDNN context model}
In this paper, we predict multiple voxels in parallel. As mentioned above, this calls for relaxing some dependencies between voxels. 
\par First, we divide a voxel block into $G$ separate groups and use $v^g$ to represent all voxels in group $g$, $g=1,\ldots,G$. We factorize the joint distribution $p(v)$ as a product of $G$ conditional distributions $p(v^g|v^{g-1},v^{g-2},\ldots,v^{1})$: 
\begin{equation}
    p(v)= \prod_{g=2}^G p(v^g|v^{g-1},v^{g-2},\ldots,v^{1}) \times p(v^1|v_{LS}).
    \label{eq:p(v)}
\end{equation}
Each term $p(v^g|v^{g-1},v^{g-2},\ldots,v^{1})$ is the joint probability of all voxels in $v^g$ being occupied given all previous groups.
Compared to VoxelDNN, we have removed the dependencies of voxels within each group. In return, we are able to predict all voxels in group $g$ in parallel. In addition, given the first group, all other groups can be autoregressively predicted. We model voxels in the first group $v^1$ as conditionally independent given the lower resolution $v_{LS}$. This procedure is applied recursively to lower resolutions until the lowest scale, which is encoded using VoxelDNN. Figure \ref{fig:msvoxeldnnarchitecture} shows the general scheme of our Multiscale VoxelDNN encoder (MSVoxelDNN). At each step of the pyramid, downsampling is obtained by applying a maxpooling operation of size $2 \times 2 \times 2$ to the high resolution block, i.e., the resulting lower resolution voxel occupancy is one if at least one of the 8 higher resolution voxels is occupied. Therefore, by training the context model to predict the first group from $v_{LS}$, we somehow learn an inverse max pooling mapping for occupancy probabilities. 



\begin{table}[t]
\caption{Number of blocks in training sets for each block size.}
\centering
\resizebox{0.97\linewidth}{!}{ \begin{tabular}{rR{0.8cm}R{0.8cm}R{0.8cm}R{1.2cm}R{1cm}}
\toprule
\begin{bf} Block size \end{bf}
&\begin{bf}MVUB\end{bf}
& \begin{bf}8i\end{bf} 
&\begin{bf}CAT1\end{bf}
& \begin{bf}ModelNet40\end{bf} 
& \begin{bf}Total\end{bf} \\
\midrule

64& 5,777&4,797 &2,777 & 1,1147&24,498\\
32& 22,082&20,436 &15,243 & 50,611&108,372\\
16& 87,578&86,106 &45,626 & 224,951&444,261\\
8& 354,617&349,760 &180,037 & 986,253&1,870,667\\
\bottomrule
\end{tabular}}
\label{table:noblocks}
\end{table}
\begin{center}
\centering
\begin{table*}[t]
\caption{Average rate in bpov of MSVoxelDNN compared with MPEG G-PCC v12 and VoxelDNN. The last column shows the gain reduction of MSVoxelDNN from VoxelDNN over G-PCC.}
\resizebox{0.99\linewidth}{!}{
\begin{tabular}{llR{1.2cm}R{1.2cm}R{2.1cm}R{1.2cm}R{2.1cm}R{1.8cm}}
\toprule
&& \multicolumn{1}{c}{\begin{bf} G-PCC \end{bf}}
& \multicolumn{2}{c}{\begin{bf} VoxelDNN \end{bf}}
& \multicolumn{3}{c}{\begin{bf} MSVoxelDNN\end{bf}} \\
\multicolumn{2}{l}{Point Cloud} &bpov&bpov&Gain over G-PCC &bpov&Gain over G-PCC & Rate increase over VoxelDNN\\
\midrule
\multicolumn{2}{l}{Microsoft \cite{loop2016microsoft}} &&&&& \\
& Phil10 &1.15 & 0.82& -29.37\% & 1.02&-11.13\% &+18.25\%\\
& Ricardo10 &1.07 & 0.74&-30.28\% & 0.95&-11.21\% &+19.07\% \\
& \textbf{Average}  &\textbf{1.11}  &\textbf{0.78}&\textbf{-28.90\%} &\textbf{0.99 } & \textbf{-11.17\%} & \textbf{+17.73\% }\\
\midrule
\multicolumn{2}{l}{MPEG \cite{noauthor_common_nodate, d20178i}} &&&&& \\
& Redandblack10  & 1.09& 0.71& -34.31\%& 0.87&-20.18\% & +14.13\%\\
& Loot10  &0.95 & 0.62& -34.16\%& 0.63&-21.05\%& +13.11\%  \\
& Thaidancer 10 &1.00 & 0.73&-27.00\% & 0.85&-15.00\% &+12.00\% \\
& Boxer 10 & 0.90&0.59 & -34.44\%&0.70 &-26.32\% & +8.12\%\\
& \textbf{Average} &\textbf{1.00} &\textbf{0.67} &\textbf{-31.79\%} &\textbf{0.79} & \textbf{-20.55\%  }&\textbf{+11.24\%} \\
\bottomrule
\end{tabular}}
\label{table:results}

\end{table*}
\end{center}

At a given scale, voxel groups are obtained by dividing the voxel block into non-overlapping $2 \times 2 \times 2$ blocks. We then select one of the 8 corners for each of $2 \times 2 \times 2$ blocks to get 8 groups. We build different models for different group predictions. Figure \ref{fig:groupdependencies} shows a grouping example and prediction scheme for group 1, 2 and 5. The other groups are modeled from previous groups in a similar manner as group 2, 5.

\subsection{Network architecture}
We employ a network structure similar to~\cite{reed2017parallel}. The network is composed of two stages: first, for each group prediction, we extract features using ResNet blocks. Compared to \cite{reed2017parallel}, we reduce the complexity of the feature extraction layer by just using 4 Resnet blocks instead of 12. The features enable to smooth out the discontinuities in the input voxel data, due to the sampling introduced with the grouping. The so-obtained spatial feature map is then partitioned into contiguous patches, such that there are $P$ patches for each dimension, and thus $P\times P\times P$ in total (we omit here for simplicity the number of channels in the feature space). In the second stage, each patch is input to a shallow auto-regressive model (in this case, a VoxelCNN). We can accelerate training/inference with parallel patches prediction instead of prediction on the whole spatial feature map (i.e., using $P>1$). However, too small patches can lead to inaccurate probabilities due to limited contexts. Therefore, we use $P=2$ which require $2^3$ small network evaluations in each forward pass ($P=4$ in \cite{reed2017parallel}).

Figure \ref{fig:networkarchitecture} shows the network architecture to predict group 2. Given group 1 of size $D\times D \times D$, the network outputs the predicted occupancy probabilities of all voxels in group 2. First, we use 4 ResNet blocks to extract a feature map from input, in each ResNet block, a 3D convolution with $3\times 3 \times 3$ filter size is placed between two $1\times 1 \times 1$ convolution layers. Next, the feature map are spatially divided into 8 patches of the same size and parallelly processed by a shallow VoxelCNN to produce occupancy probabilities. The probabilities are then merged back to a block of size $D \times D \times D \times 1$ which is the size of group 2. The shallow VoxelCNN is composed by one 3D convolutional layer with type A mask, a Resnet block followed by a 3D convolution layer. In each scale, we performs one network evaluation per group and then merge all 8 group probabilities into their spatial position in the output block.

\par Our predicted probabilities are fed as input to an arithmetic coder for lossless coding. Therefore, to minimize the output bitrate, we train MSVoxelDNN using cross-entropy loss, which is a measurement of the bitrate cost to be paid when the approximate symbol distribution $\hat{p}$ is used instead of the true symbol distribution $p$.


\subsection{Complexity analysis}\label{ssec:complexity}

The bottleneck of VoxelDNN comes from the fact that it is necessary to apply the network on each new voxel to encode/decode. If there are $d^3$ voxels in a block, VoxelDNN requires $O(d^3)$ network evaluations to estimate probabilities during decoding. For VoxelDNN encoding, it is possible to partially parallelize the process by evaluating several contexts in parallel (since they are known at the encoder side), and thus divide the computational time by a constant factor. However, this does not influence the computational complexity.

Instead, MSVoxelDNN enables to reduce substantially the computational complexity compared to VoxelDNN. At the lowest resolution the block is coded using VoxelDNN with a small number of voxels ($d=8$). Then, for each resolution level, the network is evaluated only $G$ times, where $G$ (the number of groups) is constant across scales. As we use a $2 \times 2 \times 2$ max pooling as downsampling operator in our work, the total number of levels is $\lceil{\log_8 d^3}\rceil$, and thus the complexity is $O(\log d)$.

\label{method}

\section{Experimental results }
\subsection{Experimental Setup}
\textbf{Training dataset:} We consider point clouds from different and varied datasets, including ModelNet40 \cite{wu20153d} which contains 12,311 models from 40 categories and three smaller datasets: MVUB \cite{loop2016microsoft}, MPEG CAT1 \cite{noauthor_common_nodate} and 8i \cite{d20178i}. We uniformly sample points from the mesh models from ModelNet40 and then scale them to voxelized point clouds with 9 bit precision. To enforce the fairness between the smaller datasets in which we select point clouds for testing, point clouds from MPEG CAT1 are sampled to 10 bit precision as in MVUB and 8i. 
\par To train a MSVoxelDNN model of at scale $d$ we divide all selected PCs into occupied blocks of size $d\times d\times d$. Table \ref{table:noblocks} reports the number of blocks from each dataset for training, with the majority coming from the ModelNet40 dataset. Block 8 dataset is also used to train VoxelDNN model.

\textbf{Training:} We have 3 scales and at each scale we have 8 models for 8 groups and thus, in total we train 24 MSVoxelDNN models. The mini-batch sizes are 32 at scale 64 and 64 at other scale. Our models are implemented in PyTorch and trained with Adam optimizer, a learning rate of $1e-5$ for 100 epochs on a GeForce RTX 2080 GPU.\footnote{The source code, as well as the trained models, will be made publicly available upon acceptance of the paper.}

\textbf{Experiments: } We evaluate the performance of MSVoxelDNN on a set of dense point clouds from MPEG and Microsoft datasets. These PCs were not used during training. The final bitstream is composed of the bits at all scales and the bits for the high-level octree. The average bits per occupied voxel (in ($bpov$)) is then measured by dividing the total bits by the number of occupied voxels. We compare the performance of MSVoxelDNN, VoxelDNN and G-PCC (version 12). Note that in the VoxelDNN paper \cite{nguyen2020learning}, we use a single model for all block sizes. However, in this paper, we train separate VoxelDNN models for each block sizes on the same dataset as MSVoxelDNN to have a fair comparison.

\subsection{Experimental results}

In all experiments, the high-level octree are directly converted to bytes without any compression, this part only accounts for less than $1\%$ of the bitstream.

\textbf{Rate comparison:} Table \ref{table:results} reports the rate in $bpov$ of the proposed method, MSVoxelDNN, compared with G-PCC and VoxelDNN.  We observe that MSVoxelDNN outperforms G-PCC on all test point clouds with rate savings from $11.13\%$ to $26.32\%$. Compared to VoxelDNN, MSVoxelDNN has smaller gains over G-PCC, with a bitrate increase of $8.12\%$ to $18.21\%$. This is due to the fact that MSVoxelDNN breaks some dependencies between voxels to model voxel probabilities in parallel, resulting in a less accurate context model.

\textbf{Complexity comparison:} Table \ref{table:complexity} shows the encoding/decoding run-time for G-PCC, VoxelDNN and MSVoxelDNN. It can be seen that both VoxelDNN and MSVoxelDNN are slower than G-PCC, however there is a very large speedup of MSVoxelDNN compared to VoxelDNN.  Specifically, MSVoxelDNN has a 35 and 109 times faster encoding and decoding time, respectively, compared to  VoxelDNN. The asymmetry of coding run-time is due to the possibility to partially parallelize VoxelDNN at the encoder, as mentioned in Section~\ref{ssec:complexity}.








\begin{center}
\centering
\begin{table}[ht]
\caption{Encoding/decoding time comparison per dataset (in seconds).}
\resizebox{0.97\linewidth}{!}{
\begin{tabular}{llR{1.2cm}R{1.2cm}R{1.2cm}}
\toprule

&& \multicolumn{1}{c}{\begin{bf} G-PCC \end{bf}}
& \multicolumn{1}{c}{\begin{bf} VoxelDNN \end{bf}}
& \multicolumn{1}{c}{\begin{bf} MSVoxelDNN\end{bf}} \\
\midrule
\multicolumn{2}{l}{\textbf{Encoding}} &&& \\
& Microsoft &7 & 4,124& 85  \\

& MPEG  & 3& 2,459&54  \\

\multicolumn{2}{l}{\textbf{Decoding}}&&&\\
& Microsoft &5 & 10,332 & 92  \\

& MPEG  & 3& 6,274 & 58  \\

\bottomrule
\end{tabular}}
\label{table:complexity}

\end{table}
\end{center}
\label{experiment}

\section{Conclusions}
\par In this paper, we propose a Multiscale VoxelDNN method to lossless code the geometry of dense point clouds. On this kind of content, MSVoxelDNN reduces the bitrate compared to G-PCC by up to 17\% on average, while reducing by over two orders of magnitudes the decoding complexity of the state-of-the-art VoxelDNN lossless codec. This is obtained by removing some dependencies between voxels in the same group in order to process them in parallel. 

The performance of MSVoxelDNN could be further improved by optimizing the grouping of voxels, in such a way to remove only those dependencies that do not contribute significantly to the estimation of conditional occupancy probabilities. Also, the MSVoxelDNN scheme (but this is a common issue of VoxelDNN as well) yield poor performance on sparse point clouds -- in general MSVoxelDNN has higher bitrates than G-PCC on point clouds which are not sufficiently dense. This is due to some basic hypotheses behind voxelization and convolutional neural networks, which require some substantial change of network architectures and PC representation. We are currently working towards an efficient learning-based lossless coding scheme for sparser point clouds to overcome these limitations.
\label{conclude}

\footnotesize
\bibliographystyle{IEEEtran}
\bibliography{refs}

\end{document}